\documentclass[final,5p,times,english,twocolumn]{elsarticle}

\usepackage{amssymb}
\usepackage{tensor}

\usepackage{amsmath}
\usepackage{xcolor}
\usepackage{subcaption}
\usepackage{dsfont}
\usepackage{mhchem}
\usepackage{background}

\journal{arXiv}

\usepackage{hyperref}
\hypersetup{
     colorlinks=true,
      linkcolor=blue,
}

\DeclareMathOperator{\arsinh}{arsinh}
\DeclareMathOperator{\erf}{erf}

\backgroundsetup{contents=}
\begin{document}

\begin{frontmatter}



\title{Numerical validation of a volume heated mixed fuel concept}


\author{Hartmut Ruhl and Georg Korn}

\address{Marvel Fusion, Theresienh\"ohe 12, 80339 Munich, Germany}

\begin{abstract}
While the underlying physics of the ICF approach to nuclear fusion is
well understood and a technological implementation of the indirect
drive variant of the ICF paradigm has recently been given at NIF
commercially viable ICF concepts for energy production and beyond are
still under investigation. In the present paper we propose core
elements of a novel fast direct drive mixed fuel ICF concept
that might be commercially viable. It makes use of ultra-short,
ultra-intense laser pulses interacting with nano-structured
accelerators embedded into the mixed fuel context.
The embedded accelerator technology promises to be highly efficient
and capable of fast fuel heating without fuel pre-compression
but is not the focus of the paper. It is the predominant purpose of
the mixed fuel concept to avoid cryogenic fuels since specific
chemical compounds exist that are capable of chemically binding
$\text{DT}$. To which extent mixed fuel concepts can work is
investigated in the paper. Under the assumption that the proposed
direct drive fast heating concept is capable of rapidly heating
the fuel uniformly to sufficiently high temperatures it is found with
the help of MULTI, an ICF community code, that a $\text{pBDT}$ mixed
fuel design can reach a target yield $Q_T >1$ with $\text{MJ}$
level external isochoric heating. The simulations are used to validate
a theoretical scaling model of the mixed fuel reactive hydro
flows. The paper does not present a reactor point design.
\end{abstract}

\begin{keyword}
MULTI simulations, nuclear fusion, fast
preheating, embedded nano-structured accelerators, high intensity
laser arrays.



\end{keyword}

\end{frontmatter}


\tableofcontents

\section{Introduction}\label{intro}
The indirect drive ICF approach to nuclear fusion has recently
achieved a milestone at LLNL \cite{banks2021significant,Zylstra2022,
Abu-Shawareb2022,Kritcher2022,Zylstra2022b,DOE2022,
LLNL2022,NYT2022,Science2022} demonstrating the principal viability
of inertial confinement fusion for energy production.

The implementation of the ICF concept at LLNL is an indirect drive
variant of the latter and is not considered for commercial energy
production in the community \cite{hurricane2022} raising the question
if there are alternative approaches to nuclear fusion, that might be
commercially viable.

For commercial energy production direct drive concepts are
discussed in the ICF community. Marvel Fusion investigates two
different direct drive ICF concepts. The first one limited to $\ce{DT}$
is the mainstream approach and relies on cryogenic fuels,
fuel pre-compression and subsequent hotspot generation. It is the
well-documented classic. The problem with the approach is a lack of
efficiency, a range of parasitic instabilities, and the need of
expensive cryogenic fuel technology. The second is novel and the
focus of Marvel Fusion at present. It makes use of mixed fuels without
pre-compression, does not require cryogenic fuels, and promises to be
efficient due to a proposed novel fast direct drive fuel heating
technology embedded into the fuel context consisting of
nano-structured accelerators interacting with ultra-short,
ultra-intense laser pulses. The embedded accelerators are assumed to
be capable of suppressing parasitic parametric instabilities. They
comprise both the ion beam \cite{roth2001fast} and the electron beam
\cite{tabak1994ignition} based fast fuel heating concepts. The absence
of fuel pre-compression is important since it is the prerequisite for
fast energy deposition in the fuel. The fuel is assumed to
be immobile in configuration space during energy deposition while it
almost instantaneously rearranges its momentum space.

\begin{figure}[ht]
\begin{center}
\includegraphics[width=\linewidth]{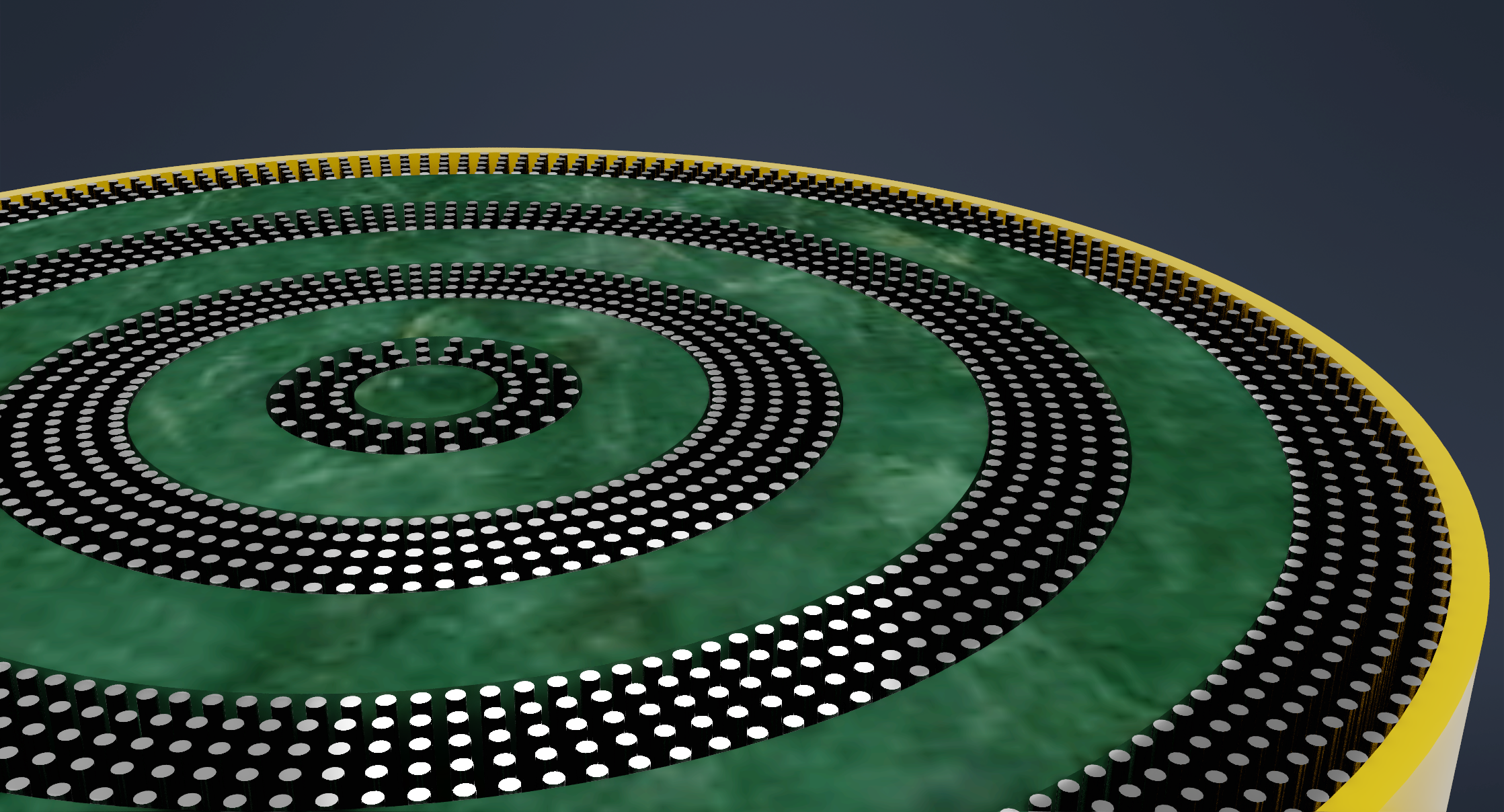}\\  
\end{center}
\caption{\label{MULTI_cylinder} Top view of a section of the reactor
  layout for the MULTI simulations. We assume an appropriate $\ce{pBDT}$ fuel 
  composite represented by the green patches in the figure. The fuel is
  interlaced with embedded accelerators illustrated by the metallic
  structures consisting of $\ce{pB}$ in the figure. The strctures are irradiated by
  ultra-short, ultra-intense laser pulses from the top and the
  bottom. The fuel is enclosed by a gold layer and heated uniformly to
  $kT_i = kT_e = 5 - 30 \,  \text{keV}$ with the help of the
  laser-irradiated embedded nano-structures. We assume that $R = L =
  1.0 - 2.0 \, \text{mm}$ hold, where $R$ is the fuel radius and $L$ its length. The fuel
  densities are $0.6 \, n_p = 0.6 \, n_D = 0.6 \, n_T =n_B$ with
  $\rho_{DT} \approx 500 \, \text{kgm}^{-3}$ and $\rho_{pB} \approx
  800 \, \text{kgm}^{-3}$. The thickness of the gold enclosure ranges
  between $0 - 1 \, \text{mm}$. The reactor sketch given here is not a
  point design.}
\end{figure}

For a sketch of the embedded accelerator concept see
Fig. \ref{MULTI_cylinder}. The metallic structures in the figure are the
nanorods embedded into the fuel. They are assumed to consist of
$\ce{pB}$. The fuel is represented by the green patches between the
nanorods. The nanorods are irradiated by multiple ultra-short,
ultra-intense laser pulses from the top and the bottow. The nanorods
Coulomb explode with high efficiency when interacting with the laser
pulses, see \cite{ruhlkornarXiv1}. The fast ions and electrons emitted
by the exploding nanorods are assumed to heat the green fuel patches
with high efficiency on a time scale much shorter than the confinement
time of the fuel.

A main advantage of an efficient heating technology is the possibility
of reactor designs that can potentially operate at low $Q_T$. Low $Q_T$
requirements promise to improve the commercial viability of a reactor
concept, in case $Q_T \gg 1 $ at reduced heating requirements could be
achieved. Combining fuel preheating with in-situ fuel compression
while there is no fuel pre-compression might allow an attractive range
of novel performant reactor designs. However, they are not the focus
of the present paper. Essentially, it is suggested that the embedded
accelerators represent a novel direct drive fast fuel heating
technology.

With the help of simulations an effective rod model can be derived.
The embedded accelerators are assumed to be composed of effective rods.
The effective rods are capable of predicting the amount of laser
energy absorbed per unit length of laser propagation along the
effective rod. In addition, the effective rod model is capable of
predicting the relative shares of the laser energy converted into ions
and electrons. Assuming that there is fast deposition of the energy of
the electrons and ions obtained from the effective rods in the fuel
surrounding the latter the number of rods in the embedded accelerators
required for a desired fusion yield $Q_T$ can be
predicted. Preliminary properties of the effective rod model have been
discussed in \cite{ruhlkornarXiv1}.

In section \ref{abstraction} an analytical scaling model describing
the reactive hydrodynamical aspects of a mixed fuel reactor filled
with $\ce{pBDT}$ is given. The proposed mixed fuel concept is new and
it is a priori not clear to which extent such fuel mixes can work. The
analytical model is intended to explain the hydrodynamical properties
of the underlying mixed fuel. Boron based chemical
compounds are capable of chemically binding $\ce{DT}$, while the
boron in the present paper is involved only marginally in fusion
reactions. The predominant purpose of the boron is to avoid
cryogenic $\ce{DT}$. We explicitly do not discuss the challenging
physics of the direct drive fast heating approach in the present
paper. In section \ref{MULTI} the reactive hydro aspects of the mixed
fuel reactor concept are validated with the help of MULTI simulations
of a $\ce{pBDT}$ filled cylindrical reactor layout. MULTI is a  
validated ICF community code. We assume that the mixed fuel reactor is
heated almost instantaneously over the entire volume to high $kT_e$
and $kT_i$, while there is no initial fluid flow. This initial reactor
setup is unusual but has interesting consequences. The simulated
reactor is enclosed by gold. However, intantaneous fuel heating without
substantial fluid motion is a technological challenge. We assume that
an instantaneous heating profile is possible with the help of the
proposed direct drive fast embedded accelerator technology, the
details of which, however, are not the focus of the present paper. In
section \ref{sum} a short summary of results is given.

\section{Mixed fuel scaling model}\label{abstraction}
Since mixed fuel concepts are new it is a priori not clear to which
extent they might work. Hence, we discuss an analytical scaling model
for mixed fuels with in-situ fusion energy feedback and effective inertial
confinement in the present paper, see also \cite{ruhlkornarXiv2} and
references therein. The analytical scaling model is intended to
support the analysis of the MULTI simulations presented in section
\ref{MULTI}. It identifies the dominant fusion relevant parameters and
their impact on fusion yield in the mixed fuel context. To some extent
the model is a generalization of an earlier one by Atzeni et
al. \cite{atzeni2004physics}, but has also imprtant novel
features. Since Marvel Fusion is a commercial entity our goal are
commercially viable variants of the established
ICF concept.

We recall that any variant of the ICF concept is inevitably thermal
due to the large discrepancies between collisional and reactive
probablities in plasma. This leads to lower temperature thresholds for electrons
and ions for an ICF reactor. In addition, any variant of the ICF
concept requires inertial confinement, which can be engineered with
the help of fuel mixes and fuel enclosures, where the enclosures can
be reactive or non-reactive. For high fusion yield in-situ fusion
energy feedback for the underlying fuel concept is required, which
again depends on the fuel mix and the selected enclosure
materials. Under certain conditions in-situ fusion energy feedback can
lead to rising reactor temperatures, which is synonymous with ignition.

We assume that the proposed direct drive fast fuel heating
concept is capable of heating the fuel of the reactor to the
temperatures $kT_e$ and $kT_i$ on a time scale much shorter than the
effective confinement time of the reactor. We assume that the reactor
can be heated such that in-situ fuel compression is avoided. Under
these assumptions we find for the lower threshold of the density -
range product of for a $\ce{pBDT}$ fuel mix
\begin{eqnarray}
  \label{explain1}        
  &&\hspace{-1cm}\rho R \\
  &&\hspace{-1cm}\ge
  \frac{\left( H_1 + H_2 \right) \, Q_F - \left( A_1 H_2 + A_2 H_1
      \right) }{2 \, \left( A_1 + A_2 - Q_F \right)} \nonumber \\
  &&\hspace{-1cm}+ \sqrt{\frac{H_1 H_2 Q_F}{A_1+A_2 -Q_F} +\left(  \frac{\left( H_1 +
     H_2 \right) \, Q_F - \left( A_1 H_2 + A_2 H_1 \right)}{2 \, \left( A_1 + A_2 -
     Q_F \right)} \right)^2} \nonumber
\end{eqnarray}
 with
 \begin{eqnarray}
 \label{QF}
 Q_F
  &=&
  \frac{ A_1 \, \rho R}{H_{1} + \rho R} + \frac{A_2 \,  \rho R}{H_{2} + \rho R} \le A_1 + A_2 \, ,
\end{eqnarray}
where
\begin{eqnarray}
A_1
&=&
\frac{2 \, \epsilon^{DT}_f \, n_D}{3 \, kT_i \, \left( n_p + n_D + n_T +
    n_B \right)} \, , \\
A_2
&=&
\frac{2 \, \epsilon^{pB}_f \, n_p}{3 \, kT_i \, \left( n_p+ n_D + n_T
    + n_B \right)}
\end{eqnarray}
and
\begin{eqnarray}
\label{HH1}
H_{1}
&\approx&
          \frac{4 \, m_p \, u^s}{\sigma^{DT}_{R0} \, u_{DT}}  \, ,  \quad
H_{2}
\approx
          \frac{4 \, m_p \, u^s}{\sigma^{pB}_{R0} \, u_{pB}} \, , \\
\label{us}
u^s &\approx& \sqrt{\frac{3 \, kT_i}{m_Z}} \, .
\end{eqnarray}
Assuming cylindrical symmetry we have
\begin{eqnarray}
  \label{explain2}  
  &&E_i
  >
  \frac{3 \pi \, kT_i \, L}{m_p \, \rho_p} \, \left( \rho R \right)^2 \, , \\
  \label{explain3}
  &&\Delta \tau 
  > \frac{1}{4 u^s \rho_p} \, \rho R \, .
\end{eqnarray}
It holds $Q_F=E_f/E_i$, where $E_f$ is the final energy in the
reactor and $E_i$ the deposited initial one. The parameter $\Delta
\tau$ is the effective confinement time. The parameter $Q_F$ is the fuel
yield. If the proposed direct drive fast fuel heating technology
is efficient $\eta \rightarrow 1$ is implied, where $\eta$ is the fuel
coupling efficiency. We then have for the fusion target yield
$Q_T = \eta \, Q_F \approx Q_F$, see \cite{ruhlkornarXiv2}. The
parameter $\rho_p$ is the proton mass density used for normalization,
$u^s$ is an effective flow velocity modeling the combined effective inertial
behavior of all fluid flows including the reactive or non-reactive
enclosure, the $\epsilon_f$ are the elementary fusion energies of all
fuels involved without neutrons, the $\sigma_{R0}$ are the fusion
cross sections of the fuel mix, $R$ is the reactor radius, $L$ is its
length, $\rho R$ is the density - range product normalized to
$\rho_p$, $kT_i$ is the initial ion temperature of the fuel mix, and
$kT_e$ the initial electron temperature. The equilibrium electron
temperature $kT_e$ is a function of $kT_i$, where $kT_e < kT_i$
holds. The parameters $n_B$, $n_p$, $n_D$, and $n_T$ are the number
densities of the fuels involved. As noted earlier the fuel ions in the
model (\ref{explain1}) - (\ref{explain3}) are mobile while the
dynamics of hydro observables is restricted to fluid rarefaction
\cite{atzeni2004physics} and $m_Z$ is the effective inertial mass 
accounting for the effective inertia all the fuel and enclosure
materials. The assumption made simplify the analytical scaling model
substantially and prove to be good enough.

According to (\ref{explain1}) - (\ref{explain3}) the temperatures
$kT_e$ and $kT_i$ are required. The temperatures $kT_e$
and $kT_i$, at which sufficient in-situ fusion energy feedback sets in
can be estimated by comparing leading density normalized power gain and
loss terms. Assuming that they are $P_f/\rho^2_p$, $P_{ie}/\rho^2_p$,
and $P_r/\rho^2_p$ and by equating
\begin{eqnarray}
\label{tgap1}
&&\frac{P_f}{\rho^2_p} = \frac{P_{ie}}{\rho^2_p} =
   \frac{P_{r}}{\rho^2_p} \, ,
\end{eqnarray}
where $P_{f}$ is the fusion power deposited in the fuel without
neutrons, $P_{ie}$ is the power transfer from all ions to electrons,
and $P_r$ is the radiation power from all electrons, lower equilibrium
temperature thresholds for $kT_e$ and $kT_i$ with $kT_e < kT_i$ are
obtained. For explicit definitions of $P_f$, $P_{ie}$, and $P_r$ see
\cite{ruhlkornarXiv2}. For the densities $0.6 \, n_p = 0.6 \, n_D =
0.6 \, n_T = n_B$ and the assumption that the total fusion $\alpha$-particle
energies are deposited in the fuel an intersection at $kT_i
\approx 20 \, \text{keV}$ with $kT_e< kT_i$ is obtained.

As outlined in \cite{ruhlkornarXiv2} the temperature $kT_i$ can rise
in case a lower temperature threshold exists, at which $P_f > P_{ie}$
and $P_f > P_r$ hold, and the time ordering
\begin{eqnarray}
\label{ordering}
&&\Delta \tau \gg t^{eq}_{ij}
\end{eqnarray}
is valid, where $i$ and $j$ denote all charged particle species and the
$t^{eq}_{ij}$ are the Spitzer equilibration times
\cite{spitzer2006physics} between particles $i$ and $j$ given by
\begin{eqnarray}
  \label{relax2}
  t^{eq}_{ij}
  &\approx& \frac{4 \pi \epsilon^2_0 \, m_i \, m_j}{Z^2_l Z^2_J \, q^4_j \, n_j \, \ln \Lambda 
         \left( n_i, n_j \right)} \, \left( \frac{kT_i}{m_l} +
            \frac{kT_j}{m_j} \right)^{\frac{3}{2}} \, . 
\end{eqnarray}
We introduce the temperature $kT_{\alpha}$ to distinguish the
temperature of the $\alpha$-particles from the general background. 
Making the simplifying assumption $kT_e = kT_i = kT$ we obtain for the
power transfer ratios between $\alpha$-particles and ions and
$\alpha$-particles and electrons, see \cite{ruhlkornarXiv2}
\begin{eqnarray}
  \frac{P_{\alpha i}}{P_{\alpha e}}
  &=&\frac{n_e \, m_e}{n_i \, m_i}
  \, \left(\frac{ 1+
      \frac{m_{\alpha} \, kT}{m_e \, kT_{\alpha}}}{1 +\frac{m_{\alpha} \,
      kT}{m_i \, kT_{\alpha}}} \right)^{\frac{3}{2}} \, .
\end{eqnarray}
There are a few cases that can be discriminated easily
\begin{eqnarray}
  \label{tgapa}
  \frac{m_i}{m_{\alpha}} < \frac{kT}{kT_{\alpha}}
  &\rightarrow&  \frac{P_{\alpha i}}{P_{\alpha e}}
  \approx \frac{n_e}{ n_i} \, \sqrt{\frac{m_i}{m_e}} \gg 1 \, , \\
  \label{tgapm}
  \frac{m_i}{m_{\alpha}} > \frac{kT}{kT_{\alpha}} > \frac{m_e}{m_{\alpha}}
  &\rightarrow&
  \frac{P_{\alpha i}}{P_{\alpha e}}
  \approx \frac{n_e \, m^{\frac{3}{2}}_{\alpha}}
  { n_i \, m_i \, \sqrt{m_e}} \, \left( \frac{kT}{kT_{\alpha}}
  \right)^{\frac{3}{2}} \, , \\    
  \frac{m_e}{m_{\alpha}} > \frac{kT}{kT_{\alpha}}
  \label{tgape}
  &\rightarrow&  \frac{P_{\alpha i}}{P_{\alpha e}}
  \approx \frac{n_e \, m_e}{ n_i \, m_i} \ll 1 \, .
\end{eqnarray}
For $kT \gg kT_{\alpha}$, see (\ref{tgape}), the power transfer from
$\alpha$-particles into ions is much larger than the power transfer
form $\alpha$-particles into electrons. For $kT \ll kT_{\alpha}$
predominantly the electrons are heated by the $\alpha$-particles.
The question is, of course, what a good estimate for $kT_{\alpha}$ is.
The worst case is equating $kT_{\alpha}$ to the fusion
$\alpha$-particle energies. 

To be capable of depositing their total energy in the fuel
volume the ranges of the fast ions generated by the presumed
embedded accelerators and the fusion related $\alpha$-particles should
preferentially be shorter than the extension of the fuel. Else, not
the total absorbed external laser and in-situ fusion energies are available
for fuel heating. A brief discussion of $\alpha$-particle stopping
power models for $\ce{DT}$ is found in \cite{xu2014effects}.
The ranges and velocities of various ions and fusion
$\alpha$-particles with initial energies $\epsilon_{i} = 3.5 \,
\text{MeV}$ in the $\ce{pBDT}$ mixed fuel background at $kT_e = 20 \,
\text{keV}$ based on the electronic stopping power excerted on the
ions according to \cite{basko1983stopping} are discussed in
\cite{ruhlkornarXiv3}.

How useful the analysis (\ref{tgap1}) - (\ref{tgape}) really is
can be debated. The embedded accelerators are capable of generating
high currents accompanied by strong electromagnetic fields.
Strong electromagnetic fields alter relaxation times. Also, there is
little reason to believe that distribution functions are Maxwellians
as is assumed when deriving (\ref{relax2}) or that the assumption of
cold monoenergetic fusion related $\alpha$-particles underlying
\cite{xu2014effects} is valid. While (\ref{tgap1}) - (\ref{tgape}) are
helpful to get an idea how in-situ fuel heating might work the details
of binary encounters between all particles have to be simulated with
the help of a detailed relativistic kinetic model comprising binary
correlations for more accurate predictions, see for example the MD
model discussed in \cite{ruhlkornarXiv1}.

Typical confinement times $\Delta \tau$ are in the $\text{ns}$ range, see
MULTI simulations in section \ref{MULTI}. The model (\ref{explain1}) -
(\ref{tgape}) highlights the influence of fusion relevant parameters on the
fusion gain $Q_T$ of the fuel. It is plausible that $Q_T > 1$ can
be obtained by quickly preheating the fuel to high temperatures $kT_e
> kT_i$, by confining the fuel, and by enabling sufficiently large
in-situ fusion energy feedback.

Since there is a long history of code development for
radiating reactive flows within the ICF community, e.g. see
\cite{ramis2016multi} and references therein, it does not really make
a lot of sense to engage in better analytical scaling models. The
focus should be on advanced numerical modeling. Hence, we rather
simulate the $\ce{pBDT}$ mixed fuel reactor for more accurate
predictions. However, it must be pointed out that most reactive hydro
codes in existence fall short when it comes to multiple interacting
resistive high velocity flows as they are generated by the embedded
accelerators in Fig. \ref{MULTI_cylinder}. Early elements for a more
appropriate numerical MD model have been outlined in \cite{ruhlkornarXiv1}.

In section \ref{MULTI} we focus on MULTI (a multi-physics, fully
Lagrangian radiation hydrodynamics code) simulations, see
\cite{ramis2016multi}, for reactor setups without in-situ fuel
compression and with uniform fuel preheating. With the help of the
simulations we try to consolidate our analytical scaling model for
$\ce{pBDT}$.

\section{MULTI simulations}\label{MULTI}
The intent of the numerical analysis based on MULTI is 
the consolidation of the mixed fuel reactor concept heated by
exawatt and $\text{MJ}$ level multi beam laser drives. The laser
pulses are assumed to be ultra-short and to have high contrast.
They are assumed to interact with small nanorods consisting of $\ce{pB}$ as
illustrated by the metallic structures in Fig. \ref{MULTI_cylinder}.
As discussed in \cite{ruhlkornarXiv1} a single small diameter nanorod
made of $\ce{pB}$ is capable of absorbing about $0.1 \,
\text{mJ}/\mu\text{m}$ corresponding to an absorption power capability
of approximately $30 \, \text{GW}$ and an absorption energy capability
of about $0.1 \, \text{J}$ for the single nanorod. The integrated
absorption power of many embedded accelerator units consisting of many
effective rods can be in the multi exawatt range. Further details of
the effective rod model are beyond the scope of the present paper.

We make use of the established community code MULTI to predict the
fusion yield $Q_T$ of uniformly preheated $\ce{pBDT}$ reactor
layouts. We assume that the preheating takes place on
time scales much shorter than the confinement time $\Delta
\tau$. We assume that there is no fuel pre-compression and
that the reactor is enclosed by gold layers of variable
thickness. As we will see there is mild in-situ compression due to the
implosion of the gold layers, since fusion $\alpha$-particles are
capable of penetrating and of heating the gold enclosure.

The nanorods explode as the laser pulses propagte along them
into the reactor. When the nanorods explode they generate
fast electrons and ions with very high efficiency. The fast electrons
and ions exiting the nanorods are assumed to be completely absorbed in
the nuclear fuel surrounding the nanorods leading to efficient
fast heating of the green fuel patches in Fig. \ref{MULTI_cylinder}
without triggering parasitic instabilities.

Since cylindrical geometry is assumed in the analytical scaling model
for the $\ce{pBDT}$ fuel mix we make use of cylindrical geometry in the
MULTI simulations. We assume
that the initial fuel densities in the MULTI simulations are close
to the natural densities of the constituents in the assumed fuel
composite consisting of $\ce{pBDT}$. In the MULTI code we interlace
$\ce{DT}$ shells with $\ce{pB}$ shells with high resolution. Since
MULTI is a fully Langragian code, the shells cannot pass each other
implying specifically that $\ce{DT}$ shells cannot penetrate $\ce{pB}$
ones. We expect that this restriction underestimates fusion yield in
the mixed fuel context. However, energy and momentum
can be exchanged between the $\ce{DT}$ and $\ce{pB}$ shells. MULTI is
capable of simulating hydro motion, correct $\alpha$-particle
stopping, heat conduction, radiation transport, and fuel
heating. MULTI comprises more realistic equations of state (for
further details see \cite{ramis2016multi}). The geometry
and the composition of the reactor simulated with MULTI is illustrated
in Fig. \ref{MULTI_cylinder}. The embedded accelerators in the figure
are emulated by appropriate $\ce{pB}$ fuel shells. Further parameters are
stated in the figure captions. It is assumed that the fuel mix
acquires uniform temperatures $kT_e= kT_i$ within a fraction of the
confinement times corresponding to the respective simulated reactor
layouts.

\begin{figure}[ht]
\begin{center}
\includegraphics[width=\linewidth]{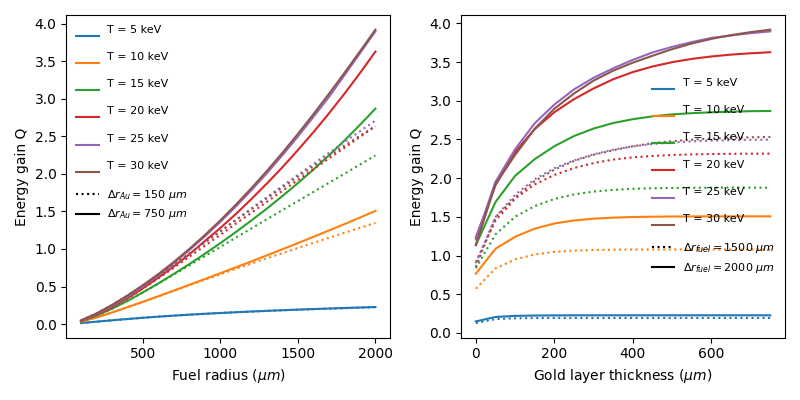}\\
\end{center}
\caption{\label{energies} MULTI simulations of fusion gains $Q_T$.
  Plot to the left:  Fusion gain vs fuel radius for the gold layer
  thicknesses $0.15 \, \text{mm}$ and $0.75 \, \text{mm}$ and a
  range of uniform initial temperatures between $0 \le kT_e= kT_i \le
  30 \, \text{keV}$. Plot to the right: Fusion gain vs gold layer
  thickness for the reactor radii $R=1.5 \, \text{mm}$ and $R=2.0 \,
  \text{mm}$. The heating energies are between $1 \, \text{MJ} < E_i <
  45 \, \text{MJ}$ depending on $R$ and the temperatures
  $kT_e=kT_i$. The fuel densities are $0.6 \, n_p = 0.6 \, n_D = 0.6
  \, n_T = n_B$ with $\rho_{DT} \approx 500 \, \text{kgm}^{-3}$ and
  $\rho_{pB} \approx 800 \, \text{kgm}^{-3}$.}
\end{figure}

A graphical illustration of the fusion gains obtained with 
the help of MULTI simulations is shown in Fig. \ref{energies}. The 
elevated inertia of the fuel due to the gold enclosure has a
significant effect as is expected by the scaling relations
(\ref{explain1}) - (\ref{relax2}). A ticker gold layer implies
increasing the effective $m_Z$ in (\ref{us}) for the fuel leading
enhanced confinement times $\Delta \tau$ and hence higher $Q_T$. As
predicted by the analytical scaling relations fusion gain also
strongly depends on the initial temperatures $kT_e= kT_i$. Higher
initial temperatures lead to higher $Q_T$. Since the reactor explodes
the initial temperatures cannot be too high.

\begin{figure}[ht]
\begin{center}
\includegraphics[width=\linewidth]{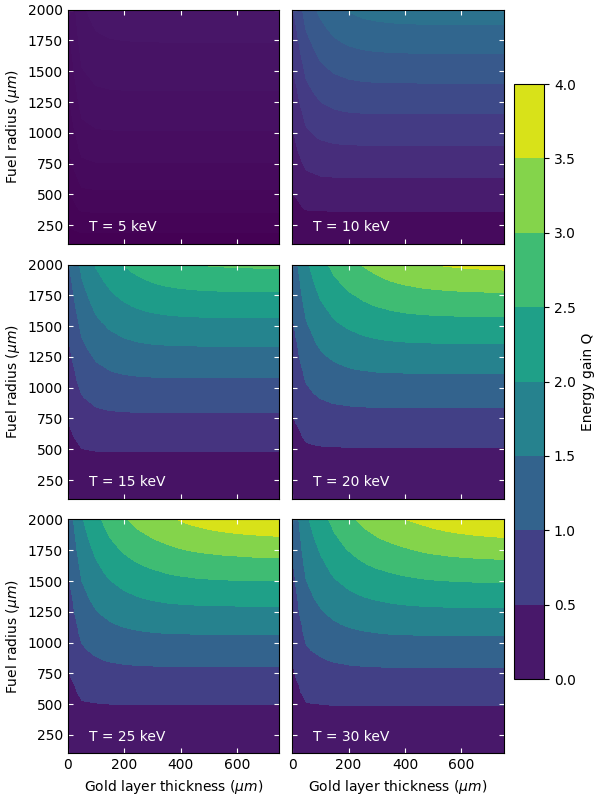}\\
\end{center}
\caption{\label{gain_overview} MULTI simulations of fusion gain
  $Q_T$ vs the reactor radii $1 \, \text{mm} \le R \le 2 \, \text{mm}$
  and gold layer thicknesses ranging between $0 - 1 \, \text{mm}$.
  The number densities are $0.6 \, n_p =  0.6 \, n_D = 0.6 \, n_T =
  n_B$, where $\rho_{DT} \approx 500 \, \text{kgm}^{-3}$ and
  $\rho_{pB}= 800 \, \text{kgm}^{-3}$. The MULTI
  simulations predict $Q_F \approx 0 - 4$ for $E_i \approx 1 - 45  \,
  \text{MJ}$.}
\end{figure}

In Fig. \ref{gain_overview} the fusion gains $Q_F$ as functions of 
reactor radius and gold layer thickness are shown for a range of 
spatially uniform initial temperatures $kT_e = kT_i$. As the plots in
Figs. \ref{energies} and \ref{gain_overview} indicate the scaling of
$Q_F$ obtained with MULTI can be approximated by a function of the
form
\begin{eqnarray}
\label{QFshape}
  Q_F 
  &=&
\frac{ A \, \rho R}{H + \rho R}
\end{eqnarray}
during the early stages of the reactors, where $A$ and $H$ are
effective parameters comprising an effective initial temperature, an
effective initial fluid density, and an initial effective fuel
mass. The shape of (\ref{QFshape}) is obtained from (\ref{QF}). The
free parameters in (\ref{QFshape}) can be obtained from the results
of the simulations shown in Fig. \ref{gain_overview}.

Since the MULTI simulations predict $E_i$ in the range of
$1 \, \text{MJ} \le E_I \le 45 \, \text{MJ}$ for gains of $0 \le Q_F
\le 4$ for the $\ce{pBDT}$ fuel mix without fuel pre-compression,
while the parametric function in (\ref{QF}) neglecting energy feedback
and enhanced fuel confinement predicts $E_i > 1 \,
\text{GJ}$ for $Q_F \approx 1$ the discrepancy to the MULTI
simulations can only be explained by some degree of in-situ energy
feedback, the increase of the effective fuel confinement in the
reactor due to the gold enclosure, and mild in-situ compression
triggered by the partial implosion of the gold layer heated by fuel ions.

\begin{figure}[ht]
\begin{center}
\includegraphics[width=\linewidth]{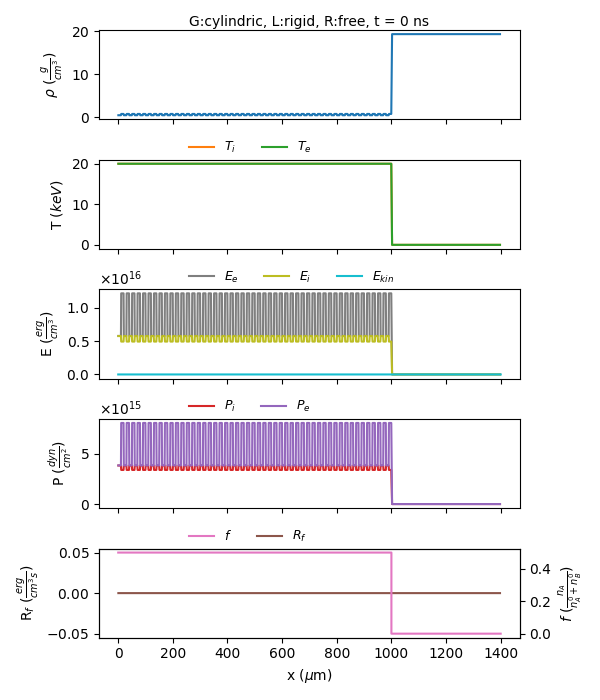}\\
\end{center}
\caption{\label{profiles000} Initial conditions of the fuel and the
  gold enclosure as a function of reactor radius for  a fuel radius of
  $R=1.1 \, \text{mm}$ with a gold layer of thickness of $0.3 \, \text{mm}$. The
  initial temperatures are $kT_e=kT_i=20 \, \text{keV}$. The number
  densities are $0.6 \, n_p =  0.6 \, n_D = 0.6 \, n_T =  n_B$, where
  $\rho_{DT} \approx 500 \, \text{kgm}^{-3}$ and $\rho_{pB} = 800 \,
  \text{kgm}^{-3}$. The quantity $f$ is the
  fraction of the burnt fuel.}
\end{figure}

\begin{figure}[ht]
\begin{center}
\includegraphics[width=\linewidth]{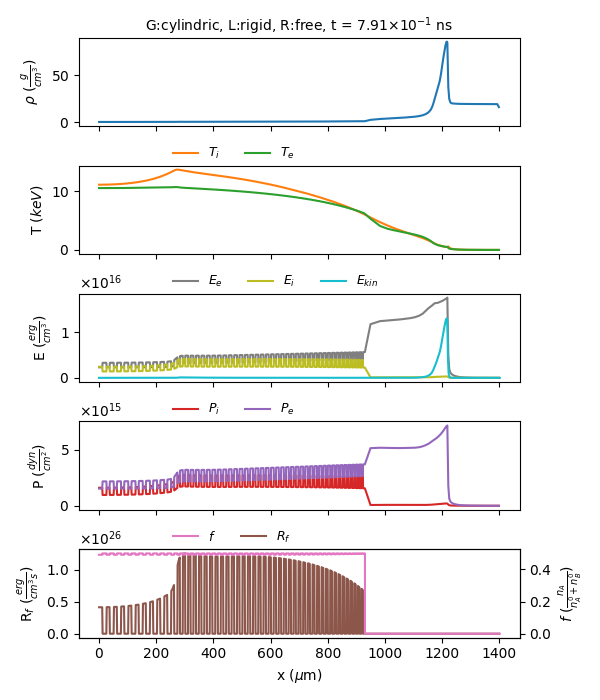}\\
\end{center}
\caption{\label{profiles200} A snapshot of a MULTI simulation at
  $t = 1 \, \text{ns}$. The initial radius is $R = 1.1 \, \text{mm}$
  and the initial gold layer thickness is $0.3 \, \text{mm}$. The
  initial temperatures are $kT_e=kT_i=20 \, \text{keV}$. The initial
  number densities are $0.6 \, n_p =  0.6 \, n_D = 0.6 \, n_T =  n_B$,
  where $\rho_{DT} = 500 \, \text{kgm}^{-3}$ and $\rho_{pB} = 800 \,
  \text{kgm}^{-3}$ is assumed. The quantity $f$
  is the fraction of burnt fuel. The simulations predict $Q_F \approx
  2$ at $E_i \approx 3.8 \, \text{MJ}$.}
\end{figure}

Figures \ref{profiles000} and \ref{profiles200} show snapshots of the
fuel and enclosure densities, the fuel and enclosure temperatures, the
fuel and enclosure energies, the fuel and enclosure pressures, and the
fuel and enclosure reaction rates vs reactor radius at $t=0$ and at $t=1 \,
\text{ns}$. Since according to the Lee and Petrasso stopping
power model, see \cite{xu2014effects}, the fusion $\alpha$-particles
lose only between $30 \% - 70 \%$ of their energy and hence are
capable of heating the gold leading to fuel temperatures dropping
below the initial ones, while $kT_e<kT_i$ holds as is expected. The
heated gold partially implodes driving mild shocks in the
fuel. Towards the end of the simulation the reactor explodes. At this
instant a large share of the energy of the system is in the gold.
The comparison of the reaction rates in Figs \ref{profiles000} and
\ref{profiles200} shows that they are higher within the fluid
shock. Another shock propagates radially out through the gold enclosure.

In section \ref{sum} we draw a few conclusions.

\section{Conclusions}\label{sum}
Marvel Fusion pursues two direct drive ICF variants. The first one is
the mainstream approach consisting of direct drive fuel
pre-compression and subsequent ion based hotspot generation. The
concept requires $\text{ns}$ low coherence laser drivers and
$\text{ps}$ laser pulses for subsequent ion beam generation. However,
direct drive fuel pre-compression is known to suffer from parasitic
parametric instabilities. Hotspot creation via laser accelerated
particle beams, see \cite{roth2001fast}, suffers from
suppressed laser energy conversion efficiency into fast ions and
potential beam transport instabilities, see \cite{ruhl2004super}, that
have to be overcome.

In contrast to the mainstream direct drive approach the fast
direct drive concept proposed by Marvel Fusion does not rely on
pre-compressed fuel, is capable of tailored fuel preheating, and
potentially in-situ fuel compression without cryogenic fuels. The
embedded accelerators, which are an integral part of the direct drive
fast heating scheme, are capable of accelerating both ions and
electrons simultaneously at extremely high efficiency, see
\cite{ruhlkornarXiv1}, both of which have been proposed as
independent fast ignition concepts earlier in the realm of ICF, see
\cite{roth2001fast,tabak1994ignition}. However, the details of the
proposed embedded accelerator based direct drive fast heating
scheme are beyond the scope of the present paper. The focus of the
present paper is on the viability of mixed fuel reactor designs
assuming the feasibility of specific fast fuel preheating
profiles in the fuel.

The analytical scaling model outlined in \cite{ruhlkornarXiv2} is
capable of predicting the correct orders of magnitude for $E_i$,
$E_f$, and $Q_T$ for a mixed fuel reactor setup consisting of
a $\ce{pBDT}$ chemical compound. According to MULTI
a fuel mix consisting of $\ce{pBDT}$ at solid density can reach $Q_F
\approx 2$ at $E_i \approx 4 \, \text{MJ}$ with the help of uniform
fuel preheating and without fuel pre-compression. In the present
paper only a specific fuel preheating profile leading to uniform
initial temperatures for $kT_e$ and $kT_i$ has been assumed.

The MULTI simulations show that mixed fuel fusion reactor layouts
with $Q_T > 1$ based on a $\ce{pBDT}$ fuel mix are conceivable if the
proposed embedded accelerator based direct drive heating technology is
available. We believe that an advanced mixed fuel fusion reactor concept
for efficient energy production without fuel pre-compression is
viable. A prerequisite, however, is a novel laser facility consisting
of many beamlines of advanced highly efficient lasers, that
can operate with $\text{ns}$ to $\text{fs}$ pulse lengths at high
rep-rate and high wall-plug efficiency yielding multiple petawatt to
multiple exawatt level laser power with $\text{MJ}$ laser pulse
energy. A laser platform providing the required technology is
available.
  
\section{Acknowledgements}
The present work has been motivated and funded by Marvel Fusion
GmbH. We would like to thank the authors of the MULTI code for
providing it to the community. Special thanks go out to Ondrej Pego
Jaura for doing the numerical modeling.

\bibliographystyle{elsarticle-num} 
\bibliography{literatur_eqn_motion}

\end{document}